\documentclass[twocolumn,showpacs,preprintnumbers,amsmath,amssymb,epsfig,widetext]{revtex4}
\usepackage{graphicx}
\usepackage{dcolumn}
\usepackage{bm}
\usepackage{epsfig}
\usepackage{color}

\def\fun#1#2{\lower3.6pt\vbox{\baselineskip0pt\lineskip.9pt
  \ialign{$\mathsurround=0pt#1\hfil##\hfil$\crcr#2\crcr\sim\crcr}}}
\def\simgt{\mathrel{\lower0.6ex\hbox{$\buildrel {\textstyle >}
 \over {\scriptstyle \sim}$}}}
\def\simlt{\mathrel{\lower0.6ex\hbox{$\buildrel {\textstyle <}
 \over {\scriptstyle \sim}$}}}

\input epsf

\newcommand{\hompc}{\,h\,{\rm Mpc}^{-1}}
\newcommand{\mpcoh}{\,h^{-1}\,{\rm Mpc}}

\newcommand{\dec}{\rm dec}

\def\be{\begin{equation}}
\def\ee{\end{equation}}
\def\ba{\begin{eqnarray}}
\def\ea{\end{eqnarray}}

\def\nn{\nonumber}

\begin{document}

\preprint{}

\title{Direct Determination of Expansion History Using Redshift Distortions}
 
\author{Yong-Seon Song}
\affiliation{Korea Astronomy and Space Science Institute, Daejeon 305-348, R. Korea}

\date{\today}

\begin{abstract}
We investigate the direct determination of expansion history using redshift distortions without plugging into detailed cosmological parameters. The observed spectra in redshift space include a mixture of information: fluctuations of density--density and velocity--velocity spectra, and distance measures of perpendicular and parallel components to the line of sight. Unfortunately it is hard to measure all the components simultaneously without any specific prior assumption. Common prior assumptions include a linear/quasi-linear model of redshift distortions or a model for the shape of the power spectra, which eventually breaks down on small scales at later epochs where nonlinear structure formation disturbs coherent growth. The degeneracy breaking, between the effect of cosmic distances and redshift distortions for example, depends on the prior we assume. An alternative approach is to utilize the cosmological principle inscribed in the heart of the Friedmann--Lema"tre--Robertson--Walker (hereafter FLRW) universe, that is, the specific relation between the angular diameter distance and the Hubble parameter, in this degeneracy breaking. We show that utilizing this FLRW prior early in the step of distinguishing the distance effect from redshift distortions helps us improve the detectability of power spectra and distance measures with no leaning on a combination of other experiments.
\end{abstract}

\pacs{draft}


\maketitle

\section{introduction}

The fundamental cosmological observables of the universe are cosmic expansion, large scale structure formation and curvature of space. Those are key tracers of the past, present and future of the universe. Cosmic expansion is defined by the expansion of the distance among cosmological objects of the universe with time. The history of the universe is known by this expansion rate at each epoch~\cite{Friedman}. The universe at large scale appears as a compilation of swiss cheese--like bubbles bordered by filaments of galaxies. This contemporary spatial distribution is clearly observable in a three dimensional reconstruction of the spectroscopy wide--deep survey~\cite{Harrison:1969fb}. The fate of the universe is determined by the global shape of space. The spatial manifold of the universe appears to be compact or non--compact without boundary~\cite{Einstein:1932zz}. The consistency of those key observables with predictions of known physical sciences on the earth will be evidence of the existence of the Friedmann--Lema"tre--Robertson--Walker (hereafter FLRW) universe. 

The discovery of metric expansion revolutionized our understanding of the universe. The metric expansion is successfully modeled by the FLRW model based upon the cosmological principle. The cosmological principle is a philosophical statement that all properties of the universe are viewed the same for all observers on a sufficiently large scale. However, the first evidence of cosmic acceleration in 1998~\cite{Riess:1998cb,Perlmutter:1998np} brought an issue of inconsistency between the observables of the FLRW universe and the predictions of physical sciences. A prime goal of precision cosmology in the next decades is to provide cosmological observables in a theoretical model independent way for a fair judgement of confirmation or exclusion. In other words, cosmic observation should be unplugged from our prior knowledge of underlying sciences. We investigate methods to probe the fundamental cosmological observables without plugging into theoretical models described by detailed cosmological parameters. 

The full history of cosmic expansion can be reconstructed using galaxy redshift surveys. Despite the enriched nonlinear structures, the zero-th order description of our current universe is homogeneous and isotropic over sufficiently large scales~\cite{York:2000gk,Peacock:2001gs,Hawkins:2002sg,Percival:2004fs,Zehavi:2004ii,Tegmark:2006az,Guzzo:2008ac,Drinkwater:2009sd,Kazin:2009cj,Reid:2009xm,Reid:2012sw}. The measured spatial distribution of galaxies  is determined by the density fluctuations and the coherent peculiar velocities of galaxies. Even though we expect the clustering of galaxies in real space to have no preferred direction, galaxy maps produced by estimating distances from redshifts obtained in spectroscopic surveys reveal an anisotropic galaxy distribution. The anisotropies arise because galaxy recession velocities, from which distances are inferred, include components from both the Hubble flow and peculiar velocities driven by the clustering of matter. Measurements of the anisotropies allow constraints to be placed on the rate of growth of clustering and Hubble flow along the line of sight~\cite{Jain:2007yk,Nesseris:2007pa,Song:2008vm,Song:2008qt,Wang:2007ht,Percival:2008sh,White:2008jy,McDonald:2008sh,Song:2010bk,Tang:2011qj,Jeong:2006xd,Jeong:2008rj,Desjacques:2009kt,Taruya:2010mx,Jennings:2010uv,Reid:2011ar,Okumura:2011pb,Kwan:2011hr,Samushia:2011cs,Blake:2011rj,Zhang:2012yt,Gaztanaga:2008xz}.

The observed spectra in redshift space depend not only on fluctuations of density and velocity fields but also on distance measures of components perpendicular and parallel to the line of sight~\cite{Blake:2003rh,Seo:2003pu,Wang:2006qt,Song:2010kq,Chuang:2012ad,Beutler:2012px}. Unfortunately, those are not simultaneously decomposed out of redshift distortions due to high degeneracy among observables. Parameterized approach based upon a specific theoretical model is commonly used to get over this problem. But those observables should be provided a priori without the theoretical model to be tested. We discuss the method to provide the following key observables in theoretical model independent way; the expansion rate at each targeted redshift, the spectra of density and peculiar velocity, and the curvature of the space. 

Here we propose a couple of theoretical model independent approaches to break the degeneracy among observables. The first approach is an easier way but less model--independent, and the second approach is a harder way but more model--independent.

First approach; we can, in principle, resolve this problem if we understand the shape of the power spectrum precisely. In the context of standard cosmology, the shape of spectra is determined before the last scattering surface, and in linear theory, it evolves coherently through all scales. In this case, the shape of spectra is determined by CMB experiments, both the coherent growth functions of density and velocity and the distance measures can be determined separately in precision using the Alcock-Paczynski test~\cite{Alcock:1979mp}. Unfortunately, this ansatz is not applicable for a specific theoretical model of cosmic acceleration in which structure formation does not grow coherently at later epochs, i.e. the determined shape of spectra has been altered since the last scattering surface~\cite{Carroll:2003wy,Dvali:2000rv,Song:2006ej}.  

Second approach; we propose an alternative method to make measurements while assuming theoretical prior as minimally as possible. In the FLRW universe, the transverse and the radial components can be unified based upon the spatial homogeneity. Instead of making additional theoretical antsatz, we propose the configuration of spectroscopic wide--deep field survey to be fully tomographic, i.e., composed of a series of redshift bins, over a certain range of redshift. If this observational constraint is satisfied, there is no further need for us to assume theoretical prior more than the FLRW universe. We find that both power spectra and distance measures are simultaneously measurable without any assistance from other experiments. Remarkably, the expansion history of Hubble flow is measured well within a couple of percentage precision at each redshift. 

We summarize the layout of this paper. In Section II, probing distance measures are presented. In Section III, model independent observation of spectra is presented. In Section IV, remaining issues are discussed.

\section{Determination of distance measures}

The observed galaxy--galaxy correlation  in redshift space depends on the fluctuations of density and velocity fields and the distance measures of components perpendicular and parallel to the line of sight. The transverse distance is represented by the angular diameter distance $D_A$, and the radial distance is represented by the inverse of Hubble flow $H^{-1}$ at each given redshift. The linear response of the observed power spectra to the variation of distance measures is studied in detail in this section. In the following subsections, we present various methods to probe distance measures out of the observed spectra. The future tomographic wide--deep survey from $z=0$ to 2 is assumed through this paper. The fiducial cosmology model is the $\Lambda CDM$ universe with cosmological parameters of $(\Omega_m=0.24,\, \Omega_k=0,\, h=0.73,\, A_S^2=2.41\times 10^{-9},\, n_S=0.96,\,\sigma_8=1.0)$.

\subsection{Components of observed galaxy power spectra}

An observed galaxy power spectrum $\tilde P(k,\mu,z)$ in redshift space is commonly modeled as~\cite{Kaiser:1987qv} over $k<0.1\hompc$,
\begin{eqnarray}
\tilde{P}(k,\mu,z) = P_{gg}(k,z) + 2\mu^2P_{g\Theta}(k,z) + \mu^4P_{\Theta\Theta}(k,z)\nn\,,\label{eq:P3D}
\end{eqnarray}
where the subscripts $g$ and $\Theta$ denote the inhomogeneity of galaxy number density and the divergence of peculiar velocity measured in the unit of $aH$. Power spectra $P_{gg}$, $P_{g\Theta}$ and $P_{\Theta\Theta}$ correspond to components of fluctuations of density and velocity fields. The subscript $g$ denotes perturbations of galaxy distribution. Galaxy number overdensity $\delta_g(k,z)$ is in general biased relative to mass fluctuations and assumed to be  $\delta_g(k,z)=b(k,z)\delta_m(k,z)$. The fiducial value of $b$ is assumed to be $b=1$ and scale independent in the manuscript.

The observed spectra are transformed by variation of distance measures through correspondent coordinate components in Fourier space. With the plane wave approximation, $k$ and $\mu$ are given by $k_{\perp}$ and $k_{\parallel}$ as,
\ba
k&=&\sqrt{k_{\perp}^2+k_{\parallel}^2} \nn \\
\mu&=&k_{\parallel}/k \,.
\ea
Given the observational quantities, such as  $k_{\perp}D_A$ (i.e., an angular scale) and $k_{\parallel}H^{-1}$ (i.e., a scale along redshift), different distance measures ($D_A$ and $H^{-1}$) result in a feature of power spectra in a different wavenumber $k_{\perp}$ and $k_{\parallel}$. When we have prior information on the location of a feature in $k_{\perp}$ and $k_{\parallel}$, we then can determine $D_A$ and $H^{-1}$.

\subsection{Alcock--Paczynski test using redshift distortions}

\begin{figure}[t]
\centering
\includegraphics[width=0.47\textwidth]{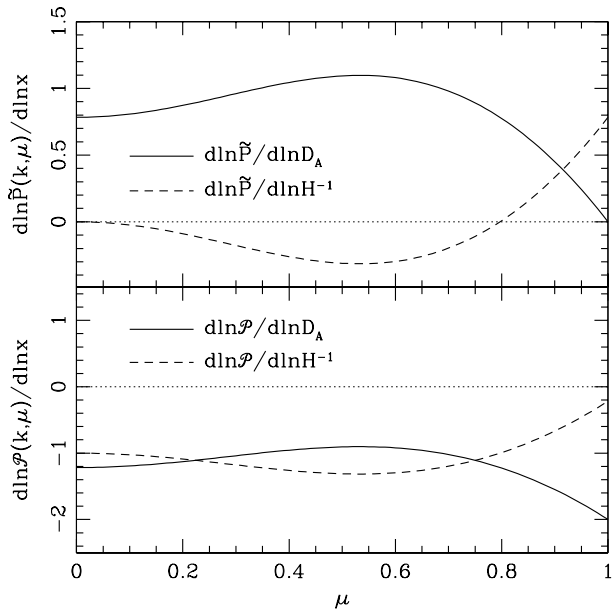}
\caption{\footnotesize ({\it Top panel}) Derivatives of $\tilde{P}$ in terms of distance measures of $D_A$ and $H^{-1}$ are presented with given $k=0.05\hompc$ and $z=1$. Solid and dash curves represent derivatives of $\tilde{P}$ in terms of $D_A$ and $H^{-1}$ respectively. ({\it Bottom panel}) Derivatives of ${\cal P}$ in terms of distance measures of $D_A$ and $H^{-1}$ are presented. Solid and dash curves represent derivatives of ${\cal P}$ in terms of $D_A$ and $H^{-1}$ respectively.}
\label{fig:dPksdx}
\end{figure}

Ratio between observed transverse and radial distances varies with the assumed theoretical models. If an object is known to be isotropic, the ratio of the intrinsic transverse and radial distances gives a relation of the observed distance measures of $D_A$ and $H^{-1}$. This method is called as Alcock--Paczynski test. The original Alcock-Paczynski test can be extended into the intrinsically anisotropic objects. If the anisotropy of the observed spectra is known, the ratio between true $D_A$ and $H^{-1}$ should be known by the anisotropic relation of the observed distance measures. 

We discuss the extended Alcock-Paczynski test using redshift distortions in this subsection. To begin with, the spectra of $P_{gg}$ and $P_{\Theta\Theta}$ are assumed to be known here. The observed anisotropy of $\tilde{P}$ is configured along the $\mu$ direction. We present the correspondent change of $\tilde{P}(k,\mu)$ to the variation of $D_A$ or $H^{-1}$.

Logarithmic derivative of $\tilde{P}$ in terms of $D_A$ is given by,
\ba\label{eq:dPksdDA}
\frac{d\ln\tilde{P}(k,\mu)}{d\ln D_A} &=& -(1-\mu^2)\frac{d\ln\tilde{P}(k,\mu)}{d\ln k}\\
&+&4\mu^2(1-\mu^2)\frac{P_{g\Theta}(k)+\mu^2P_{\Theta\Theta}(k)}{\tilde{P}(k,\mu)}\,,\nn
\ea
and the logarithmic derivative of $\tilde{P}$ in terms of $H^{-1}$ is given by,
\ba\label{eq:dPksdHin}
\frac{d\ln\tilde{P}(k,\mu)}{d\ln H^{-1}} &=& -\mu^2\frac{d\ln\tilde{P}(k,\mu)}{d\ln k}\\
&-&4\mu^2(1-\mu^2)\frac{P_{g\Theta}(k)+\mu^2P_{\Theta\Theta}(k)}{\tilde{P}(k,\mu)}\,,\nn
\ea
where $d\ln\tilde{P}/d\ln k$ is,
\ba
\frac{d\ln\tilde{P}}{d\ln k}=\frac{d\ln P_{gg}}{d\ln k}+2\mu^2\frac{d\ln P_{g\Theta}}{d\ln k}+\mu^4\frac{d\ln P_{\Theta\Theta}}{d\ln k}\,.\nn
\ea
The derivatives are presented graphically in Fig.~\ref{fig:dPksdx}. The derivative of $d\ln\tilde{P}/d\ln D_A$ vanishes at $\mu\rightarrow 1$. The orientation of correlation pairs of $\tilde{P}$ at $\mu\rightarrow 1$ is parallel to the line of sight. Thus $\tilde{P}$ becomes independent of the variation of transverse distance of $D_A$. The derivative of $d\ln\tilde{P}/d\ln H^{-1}$ vanishes at $\mu\rightarrow 0$ for the opposite reason.

We find that from redshift distortion alone, the radial and transverse correlations can be measured. The two point galaxy correlation function is determined at linear order in perturbation theory. Redshift space distortions are taken into account for arbitrary angular and redshift separations. The observed spectra of $\tilde{P}$ in redshift space are very different along the $\mu$ direction. The transverse and radial distances are indeed separable as shown in Fig.~\ref{fig:dPksdx}. It can be the extended Alcock-Paczynski test using redshift distortions. When the spectra of $P_{gg}$ and $P_{\Theta\Theta}$ are known, $D_A$ and $H^{-1}$ will be measured with good precision.

In practice, the observed spectra are estimated using referenced coordinates, not true coordinates, because the transformation rule between the observed (Ra, Dec, $z$) and the cartesian coordinates of distance measures is unknown. This uncertainty is added as a volume factor, which is given by,
\ba
\tilde{P}(k_{\rm ref},\mu_{\rm ref};z)/V_{\rm ref} = \tilde{P}(k,\mu;z)/V
\ea
where $V$ is a volume factor given by $V=D_A^2H^{-1}$. If we define projected spectra $\cal P$ as,
\ba
{\cal P}(k,\mu,z)\equiv \tilde P(k,\mu,z)\left[V_{\rm ref}(z)/V(z)\right]\,,
\ea
then the logarithmic derivatives of ${\cal P}$ in terms of $D_A$ and $H^{-1}$ are given by,
\ba\label{eq:dPksdDAV}
\frac{d\ln{\cal P}(k,\mu)}{d\ln D_A} &=& \frac{d\ln\tilde{P}(k,\mu)}{d\ln D_A} -2 
\ea
\ba\label{eq:dPksdHinV}
\frac{d\ln{\cal P}(k,\mu)}{d\ln H^{-1}} &=& \frac{d\ln\tilde{P}(k,\mu)}{d\ln H^{-1}} -1\,.
\ea
Constants in the above equations represent the volume factor effects. If those effects dominate, then the orthogonal feature of $D_A$ and $H^{-1}$ will be wiped out. We present results in the bottom panel of Fig.~\ref{fig:dPksdx}. Although it becomes degenerate at $\mu\rightarrow 0$, a distinct feature still remains at $\mu\rightarrow 1$, which enables us to separately measure $D_A$ and $H^{-1}$. Hereafter, this volume factor is included in the derivatives of $\tilde{P}$.

This Alcock-Paczynski test using redshift distortions is available, when the anisotropy is known previously. The anisotropy in the observed spectra is caused by the squeezing effect resulting from radial recessing velocities of galaxies. In reality, it is unknown. In the following subsections, we discuss the specific conditions to measure both the transverse and radial distances, while the spectra of $P_{gg}$ and $P_{\Theta\Theta}$ are not given.

\subsection{Estimation to derive detectability}

We estimate errors to decompose all those components introduced in the previous subsection using the following Fisher matrix analysis~\cite{White:2008jy}. With the consideration of the non-perturbative contribution and the perfect correlation between $g$ and $\Theta$ at linear regime, the power spectra in redshift space, $\tilde{P}$ in Eq.~\ref{eq:P3D}, are rewritten as~\cite{Scoccimarro:2004tg},
\begin{eqnarray}
\tilde{P}(k,\mu,z) &=& \big\{P_{gg}(k,z)
   + 2\mu^2r(k)\left[P_{gg}(k,z)P_{\Theta\Theta}(k,z)\right]^{1/2}\nonumber\\
   &+& \mu^4P_{\Theta\Theta}(k,z)\big\}F_{\rm FoG}(k,\mu,\sigma_z) \,.
\end{eqnarray}
The cross-correlation coefficient $r(k)$ is defined as $r(k)\equiv P_{g\Theta}/\sqrt{P_{gg}P_{\Theta\Theta}}$. The density--velocity fields are highly correlated for $k<0.1\,h\,{\rm Mpc}^{-1}$ thus we assume  that the density and velocities are perfectly correlated, $r(k)\sim 1$. The density-velocity cross-spectra become the geometric mean of the two auto-spectra to leave only two free functions, $P_{gg}$ and $P_{\Theta\Theta}$. The function of $F_{\rm FoG}$ represents the Finger-of-God effect~\cite{Fisher:1994ks,Scoccimarro:2004tg,Matsubara:2007wj,Crocce:2007dt,Taruya:2010mx}. For simplicity, we set $F_{\rm FoG}=1$ in this work.

Errors of determining parameter $p_{\alpha}$ out of $\tilde{P}$ can be estimated using Fisher matrix analysis determining the sensitivity of a particular measurement. The Fisher matrix for this decomposition, $F_{\alpha\beta}^{\rm dec}$, is written as,
\ba\label{eq:Fdec}
F_{\alpha\beta}^{\dec}=\int\frac{\partial\tilde{P}(\vec{k})}{\partial p_{\alpha}}\frac{V_{\rm eff}(\tilde{P})}{\tilde{P}(\vec{k})^2}\frac{\partial\tilde{P}(\vec{k})}{\partial p_{\beta}}\frac{d^3k}{2(2\pi)^3}\,.
\ea
The effective volume $V_{\rm eff}(\tilde{P})$ is given by,
\ba
V_{\rm eff}(\tilde{P})=\left[\frac{n\tilde{P}}{n\tilde{P}+1}\right]^2V_{\rm survey}\,,
\ea
where $n$ denotes galaxy number density, here $n=5\times 10^{-3}(\mpcoh)^{-3}$ which is an approximated average value using estimation for the future sky wide--deep field survey in~\cite{Cimatti:2009is}. Comoving volume, $V_{\rm survey}$, given by each redshift shell from $z=0$ to 2 with $\Delta z=0.2$ ($f_{sky}=1/2$) is written as, 
\ba
V_{\rm survey}=f_{sky}\frac{4\pi}{3}(D_{\rm outer}^3-D_{\rm inner}^3)\,,
\ea
where $D_{\rm outer}$ and $D_{\rm inner}$ denote comoving distances of outer and inner shell of the given redshift bin respectively. Theoretical estimation of density--density and velocity--velocity spectra using this method is studied in~\cite{White:2008jy}.

\subsection{Approach I: shape of spectra prior}\label{sec:G}

\begin{figure*}[t]
\centering
\includegraphics[width=0.47\textwidth]{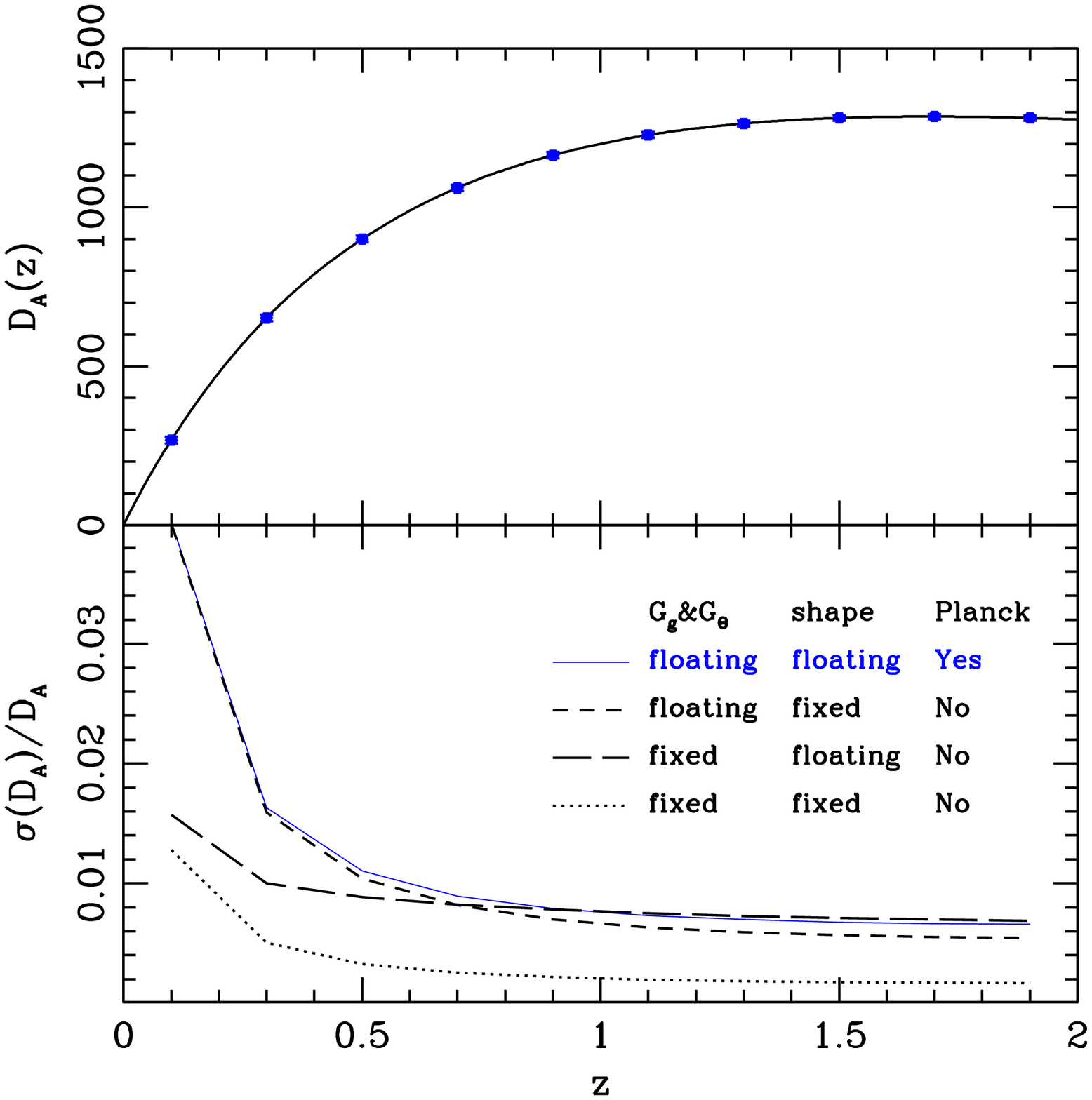}\hfill
\includegraphics[width=0.47\textwidth]{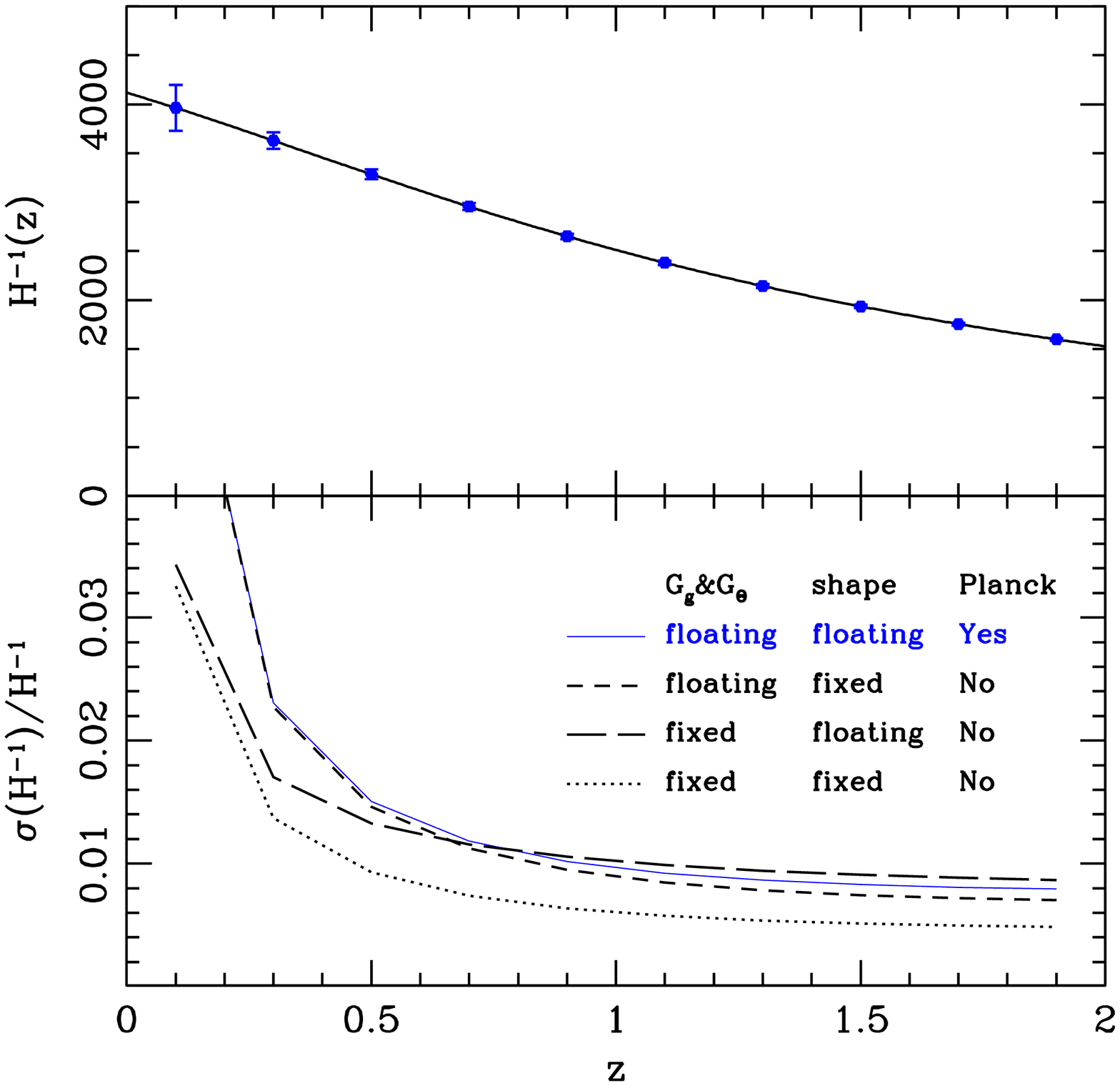}
\caption{\footnotesize {\it (Left panel)} The fiducial $D_A(z)$ is presented as black curve in the top panel. Cases for the fractional errors are presented in the bottom panel; dotted curve (fixed $G_g$ $\&$ $G_{\Theta}$, fixed shape of spectra, no Planck prior), dash curve (floating $G_g$ $\&$ $G_{\Theta}$, fixed shape of spectra, no Planck prior), long dash curve (fixed $G_g$ $\&$ $G_{\Theta}$, floating shape of spectra, no Planck prior), and thin solid curve (floating $G_g$ $\&$ $G_{\Theta}$, floating shape of spectra, applying Planck prior). {\it (Right panel)} The same captions for $H^{-1}$ as in the left panel.}
\label{fig:DAHin}
\end{figure*}

In this subsection, we present an easier approach to determine distance measures. Although the full information of spectra are unknown, $D_A$ and $H^{-1}$ are observable with the condition of the shape of spectra given. 

The shape of the power spectra of density--density and velocity--velocity correlations critically depends on the epoch of matter--radiation equality. Under the paradigm of inflationary theory, initial fluctuations are stretched outside the horizon at different epochs which generates the tilt in the power spectrum. The predicted initial tilting is parameterised as a spectral index ($n_S$) which is just the shape dependence due to the initial conditions. When the initial fluctuations reach the coherent evolution epoch after matter-radiation equality, they experience a scale-dependent shift from the moment they re-enter the horizon to the equality epoch. Gravitational instability is governed by the interplay between radiative pressure resistance and gravitational infall. The different duration of modes during this period result in a secondary shape dependence in the power spectrum. This shape dependence is determined by the ratio between matter and radiation energy densities and sets the location of the matter-radiation equality in the time coordinate~\cite{Eisenstein:1997ik}. Therefore, the broadband shape is well determined, in a model--independent way,  when the shape parameters of $n_S$, $\omega_m$ and $\omega_b$ are given~\cite{Song:2010kq}. 

Assuming that linear theory holds,
 the parameter set $p_{\alpha}$ is given by ($G_g(z_j)$, $G_{\Theta}(z_j)$, $D_A(z_j)$, $H^{-1}(z_j)$, $\omega_m$, $\omega_b$, $n_S$). Here $G_X$ denotes the coherent growth factor defined by,
\ba\label{eq:pall}
P_{\Phi\Phi}(k,a)&=&Q_{\Phi}(k)G_{\Phi}^2(a),\nn\\
P_{gg}(k,a)&=&Q_{m}(k)G_{g}^2(a),\nn\\
P_{\Theta\Theta}(k,a)&=&Q_m(k)G_{\Theta}^2(a),
\ea
where the subscript $\Phi$ denotes the curvature perturbation in the Newtonian gauge, 
\begin{equation}
 ds^2=-(1+2\Psi)dt^2+a^2(1+2\Phi)dx^2\,.
\end{equation}
The growth function $G_g$ is defined as $G_{g}\equiv b\,g_{\delta_m}$ where $b$ is the standard linear bias parameter between the density of galaxies and the underlying dark matter, $\delta_m$. As we follow the positive sign conversion, $G_{\Theta}$ is the growth function of $-\Theta$. We further assume the scale independent bias at scales $k<0.1\hompc$. The shape factor of the perturbed metric power spectra $Q_{\Phi}(k)$ is defined as
\be
Q_{\Phi}(k)=\frac{2\pi^2}{k^3}\frac{9}{25}\Delta^2_{\zeta_0}(k)T^2_{\Phi}(k),
\ee
which is a dimensionless metric power spectra at $a_{eq}$ (the matter--radiation equilibrium epoch), and $\Delta^2_{\zeta_0}(k)$ is the initial fluctuations in the comoving gauge and  $T_{\Phi}(k)$ is the transfer function normalised at $T_{\Phi}(k\rightarrow 0)=1$. The primordial shape $\Delta^2_{\zeta_0}(k)$ depends on $n_S$ (the slope of the primordial power spectrum), as $\Delta^2_{\zeta_0}(k)=A^2_S(k/k_p)^{n_S-1}$, where $A^2_S$ is the amplitude of the initial comoving fluctuations at the pivot scale, $k_p=0.002$ ${\rm Mpc}^{-1}$. The intermediate shape factor $T_{\Phi}(k)$ depends on $\omega_m$. 

The shape factor for the matter fluctuations, $Q_m(k)$, which is important for both the galaxy--galaxy and velocity--velocity power spectra in Eq. \ref{eq:pall} above, is given by the conversion from $Q_{\Phi}(k)$ of,
\be
Q_m(k)\equiv\frac{4}{9}\frac{k^4}{H_0^4\Omega_m^2}Q_{\Phi}(k),
\ee
where, assuming $c=1$, we can write $H_0\equiv 1/2997\hompc$. 

Derivatives in Fisher matrix in terms of ($G_g(z_j)$, $G_{\Theta}(z_j)$ are given by,
\begin{eqnarray}
  \frac{\partial \ln \tilde{P}(k,\mu,z_j)}{\partial G_{g}(z_j)}
  &=& \frac{2Q_m(k)}{\tilde{P}(k,\mu,z_j)}
  \left[G_{g}(z_j) + \mu^2G_{\Theta}(z_j) \right] \nonumber \\
  \frac{\partial\ln \tilde{P}(k,\mu,z_j)}{\partial G_{\Theta}(z_j)}
  &=&\frac{2\mu^2Q_m(k)}{\tilde{P}(k,\mu,z_j)}
  \left[G_{g}(z_j)+\mu^2G_{\Theta}(z_j)\right]\,,
\end{eqnarray}
and derivatives in terms of ($D_A(z_j)$, $H^{-1}(z_j)$) are given by,
\begin{widetext}
\begin{eqnarray}\label{eq:dpdD}
  \frac{\partial\ln \tilde{P}(k,\mu,z_j)}{\partial \ln D_A(z_j)}
  &=& \frac{1}{\tilde{P}(k,\mu,z_j)}\left[\frac{\partial \tilde{P}(k,\mu,z_j)}{\partial \ln k}\frac{\partial \ln k}{\partial \ln D_A(z_j)}+\frac{\partial \tilde{P}(k,\mu,z_j)}{\partial \ln \mu}\frac{\partial \ln \mu}{\partial \ln D_A(z_j)}\right]\nn \\
 \frac{\partial\ln \tilde{P}(k,\mu,z_j)}{\partial \ln H^{-1}(z_j)}
  &=& \frac{1}{\tilde{P}(k,\mu,z_j)}\left[\frac{\partial \tilde{P}(k,\mu,z_j)}{\partial \ln k}\frac{\partial \ln k}{\partial \ln H^{-1}(z_j)}+\frac{\partial \tilde{P}(k,\mu,z_j)}{\partial \ln \mu}\frac{\partial \ln \mu}{\partial \ln H^{-1}(z_j)}\right]\,,
\end{eqnarray}
\end{widetext}
where,
\begin{eqnarray}
\frac{\partial \ln k}{\partial \ln D_A}&=& -(1-\mu^2)\nn\\
\frac{\partial \ln \mu}{\partial \ln D_A}&=& (1-\mu^2)\nn \\
\frac{\partial \ln k}{\partial \ln H^{-1}}&=& -\mu^2\nn\\
\frac{\partial \ln \mu}{\partial \ln H^{-1}}&=& -(1-\mu^2)\,.
\end{eqnarray}
Fisher matrix components with the variation of ($\omega_m$, $\omega_b$, $n_S$) are calculated computationally using CAMB output.

Results are presented in Fig.~\ref{fig:DAHin}. When the spectra are known, the transverse and radial distances are measured with high precision. The dash curves in the bottom panels of Fig.~\ref{fig:DAHin} represent fractional errors, when only the overall amplitudes of $G_g$ and $G_{\Theta}$ are unknown. It shows that the Alcock-Paczynski test is still available, when the shape of spectra is given at least. However, when both the growth functions and the shape of spectra are unknown, $D_A$ and $H^{-1}$ are not separately measurable. 

In Approach I, we provide the information of the shape of spectra externally using CMB experiments. If the structure formation evolves coherently since the last scattering surface, the shape of spectra is determined by CMB precisely. We apply prior information of the shape parameters for the Fisher matrix in which both the growth functions and the shape of spectra remains unknown. The diagonal elements of ($\omega_m$, $\omega_b$, $n_S$) in the Fisher matrix are given by; $\sigma(\omega_m)=4.9\times 10^{-4}$, $\sigma(\omega_b)=3.7\times 10^{-5}$, $\sigma(n_S)=1.7\times 10^{-3}$. Here Planck experiment specs are applied. The thin blue solid curves in Fig.~\ref{fig:DAHin} represent the detectability of $D_A$ and $H^{-1}$ with this Planck prior applied. The detectability approaches to the dash curves. We proves that uncertainties caused by unknown shape of spectra are resolved by applying CMB prior. 

\subsection{Approach II: FLRW prior}

An alternative method is proposed in this subsection. Approach I is not applicable without the combination of CMB experiments. In addition, the shape of spectra is not always determined before the last scattering surface. The assumption of coherent growth of structure formation at later epoch is not valid for some theoretical models such as f(R) gravity. Then Approach I will bias observables. 

Our understanding of the universe is based upon the cosmological principle. In a homogeneous universe, two distinct components of distance measures, $D_A$ and $H^{-1}$, are dependent on each other. The angular diameter distance is an integrated sum of Hubble flow from the current to the targeted redshift. If the wide--deep field survey is designed to be fully tomographic through all redshifts, then both different components of distance measures can be unified. We investigate the detectability of Hubble flow $H^{-1}$, when distance measures are unified into a single degree of freedom. The growth functions and the shape of spectra remains unknown in this subsection.

With the suggested tomographical wide--deep field survey design, the angular diameter distance of $D_A$ is approximately expressed by the discrete sum of Hubble flow of $H^{-1}$ at each redshift bin as,
\ba\label{eq:daandH}
D_A(z_j)&=&\frac{1}{1+z_j}\int^{z_j}_0 dz' H^{-1}(z') \nn \\
&\sim& \frac{1}{1+z_j}\sum^{N_j}_{j'=1} H^{-1}(z_{j'})\Delta z_{j'}\,,
\ea
where $N_j$ represents the total number of redshift bins up to the targeted redshift of $z_j$. Then $p_{\alpha}$ parameter space for distance measures is unified into a single parameter of Hubble flow $H^{-1}$. Most future spectroscopy wide--deep field redshift survey programs such as SUBARU PFS~\cite{Ellis:2012rn}, BigBOSS~\cite{Schlegel:2011wb}, DESpec~\cite{Abdalla:2012fw} and EUCLID~\cite{Cimatti:2009is} are planned to make full tomographic scanning. The parameter set of $p_{\alpha}$ becomes ($G_g(z_j)$, $G_{\Theta}(z_j)$, $H^{-1}(z_j')$, $\omega_m$, $\omega_b$, $n_S$). The derivative in terms of $H^{-1}(z_j)$ is given by,
\begin{eqnarray}
  \frac{\partial\ln \tilde{P}(k,\mu,z_j)}{\partial \ln H^{-1}(z_{j'})}
  &=& \frac{1}{\tilde{P}(k,\mu,z_j)}\frac{\partial \tilde{P}(k,\mu,z_j)}{\partial \ln k}\frac{\partial \ln k}{\partial \ln H^{-1}(z_{j'})}\nn\\
&+&\frac{1}{\tilde{P}(k,\mu,z_j)}\frac{\partial \tilde{P}(k,\mu,z_j)}{\partial \ln \mu}\frac{\partial \ln \mu}{\partial \ln H^{-1}(z_{j'})}\nn
\end{eqnarray}
where
\begin{eqnarray}
\frac{\partial \ln k}{\partial \ln H^{-1}(z_{j'})}&=&\frac{\partial \ln k}{\partial \ln D_A(z_{j})}\frac{\partial \ln D_A(z_{j})}{\partial \ln H^{-1}(z_{j'})}-\mu^2\delta_{jj'}  \\
\frac{\partial \ln \mu}{\partial \ln H^{-1}(z_{j'})}&=&\frac{\partial \ln \mu}{\partial \ln D_A(z_{j})}\frac{\partial \ln D_A(z_{j})}{\partial \ln H^{-1}(z_{j'})}-(1-\mu^2)\delta_{jj'} \nn\,,
\end{eqnarray}
where $z_{j'}<z_{j}$ and $\delta_{jj'}=1$ at $j=j'$ otherwise 0. The reduction of parameter space for distance measures enhances the detectability of $p_{\alpha}$ parameter.

\begin{figure}[t]
\centering
\includegraphics[width=0.47\textwidth]{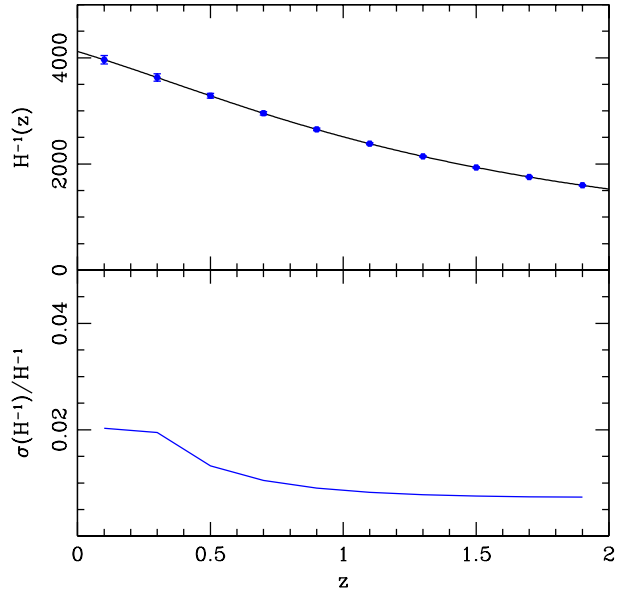}
\caption{\footnotesize Black solid curve in the top panel represents fiducial $H^{-1}(z)$ with error bars estimated using FLRW prior in small curvature approximation. Blue solid curve in the bottom panel represents fractional errors.}
\label{fig:H_FRW_with_G}
\end{figure}

Results are presented in Fig.~\ref{fig:H_FRW_with_G}. With the unknown growth functions and the shape of spectra, fractional errors of radial distance $H^{-1}$ are fairly small. We find that the fractional errors of $H^{-1}$ do not increase much from the case in which the whole spectra are assume to be known. Hubble flow is directly observable about a percentage precision through all redshifts, although the true cosmological model of the universe is not yet determined. Most Hubble flow probes are possible by underlying cosmological models such as $\Lambda CDM$ universe. But here we claim that $H^{-1}$ is directly measurable regardless of the determined cosmological model.

Additionally, it is interesting to observe that the shape parameters are self--determined with no combination of CMB experiments. The estimated constraints are;  $\sigma(\omega_m)=1.1\times 10^{-3}$, $\sigma(\omega_b)=5.8\times 10^{-4}$, $\sigma(n_S)=5.9\times 10^{-3}$. We discuss the detectability of the growth functions in the next section.

The expression of $D_A$ in terms of $H^{-1}$ in Eq.~\ref{eq:daandH} is only available in the flat universe. With the non--trivial curvature, the equation of Eq.~\ref{eq:daandH} is updated in the following way. The curvature parameter is given by,
\ba
{\cal K} = \sqrt{|\Omega_k|h^2} / 2997.2 \hompc\,.
\ea
Then parameter space of $p_{\alpha}$ is extended into ($G_g(z_j)$, $G_{\Theta}(z_j)$, $H^{-1}(z_j')$, $\omega_m$, $\omega_b$, $n_S$, ${\cal K}$). The estimated angular diameter distance $D_A(z_j)$ is given with open curvature as,
\ba
D_A(z_j) = \frac{1}{{\cal K} (1+z_j)} \sinh \left[{\cal K}\sum^{N_j}_{j'=1} H^{-1}(z_{j'})\Delta z_{j'}\right]\,,
\ea
and with closed curvature as,
\ba
D_A(z_j) = \frac{1}{{\cal K} (1+z_j)} \sin \left[{\cal K}\sum^{N_j}_{j'=1} H^{-1}(z_{j'})\Delta z_{j'}\right]\,.
\ea
We investigate the impact on the detectability of $H^{-1}$ with curvature parameter. The estimated constraints on $H^{-1}$ become poorer only by a factor of 1.5 with curvature marginalized. 

There has been no significance deviation of curvature parameter from flatness observed. Additionally, the approximate flatness is predicted by the inflationary theoretical model. Therefore small curvature is favored not only observationally but also theoretically. In this small curvature approximation of ${\cal K} \ll 1$, the expression of $D_A$ can be approximately given by Eq.~\ref{eq:daandH} at the redshift of $z<2$. Hereafter, the small curvature approximation is assumed. 

\section{Challenge to measure spectra in model independent way}

\begin{figure}[t]
\centering
\includegraphics[width=0.47\textwidth]{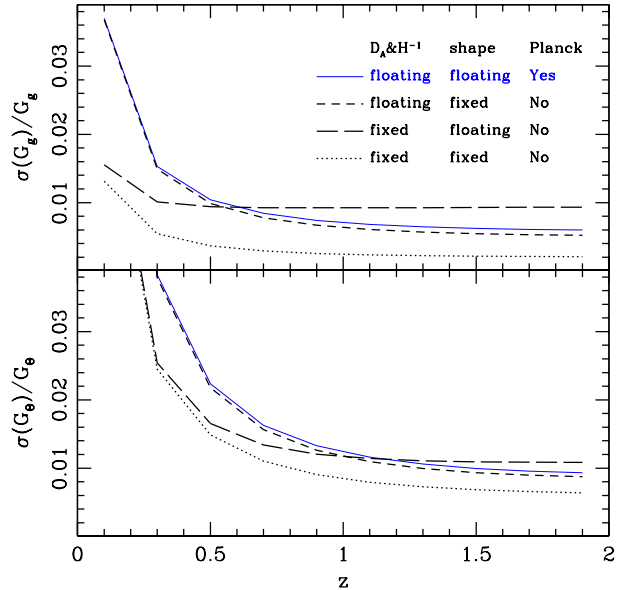}
\caption{\footnotesize {\it (Top panel)} Cases for the fractional errors are presented; dotted curve (fixed $D_A$ $\&$ $H^{-1}$, fixed shape of spectra, no Planck prior), dash curve (floating $D_A$ $\&$ $H^{-1}$, fixed shape of spectra, no Planck prior), long dash curve (fixed $D_A$ $\&$ $H^{-1}$, floating shape of spectra, no Planck prior), and thin solid curve (floating $D_A$ $\&$ $H^{-1}$, floating shape of spectra, applying Planck prior). {\it (Bottom panel)} The same captions for $G_{\Theta}$ as in the took panel.}
\label{fig:GgGT}
\end{figure}

Predictions for the error on $G_g$ and $G_{\Theta}$ are shown in Fig.~\ref{fig:GgGT}. Results are presented either assuming that the distance measures or the shape of spectra is unknown. Fractional error on $G_g$ and $G_{\Theta}$ increases by a factor of 1.5--3, when either information is not known. However, none of $G_g$ and $G_{\Theta}$ are observable when both information are not given at the same time. It is useful to ask how these forecasts depend on the input assumptions. The shape of spectra can be externally given by CMB experiments. The thin blue solid curves in Fig.~\ref{fig:GgGT} represent the detectability of $G_g$ and $G_{\Theta}$ with Planck prior applied. This Approach I turns out to be useful for probing overall amplitudes of $G_g$ and $G_{\Theta}$, while the distance measures are still unknown.

Unfortunately, Approach I is applicable for a limited number of theoretical models. If the structure formation does not grow coherently at later epoch~\cite{Carroll:2003wy,Dvali:2000rv}, the shape of spectra is not determined at the last scattering surface. Our prime interest is to probe the whole spectra represented by $P_{gg}(k_i,z_j)$ and $P_{\Theta\Theta}(k_i,z_j)$. The parametrization of $P_{gg}(k_i,z_j)$ and $P_{\Theta\Theta}(k_i,z_j)$ includes the uncertainties of both the growth functions and the shape of spectra. Therefore, if the distance measures are unknown, $P_{gg}(k_i,z_j)$ and $P_{\Theta\Theta}(k_i,z_j)$ will not be probed. In this section, we present the condition that the whole spectra of $P_{gg}(k_i,z_j)$ and $P_{\Theta\Theta}(k_i,z_j)$ are observable using Approach II.

\begin{figure*}[t]
\centering
\includegraphics[width=0.47\textwidth]{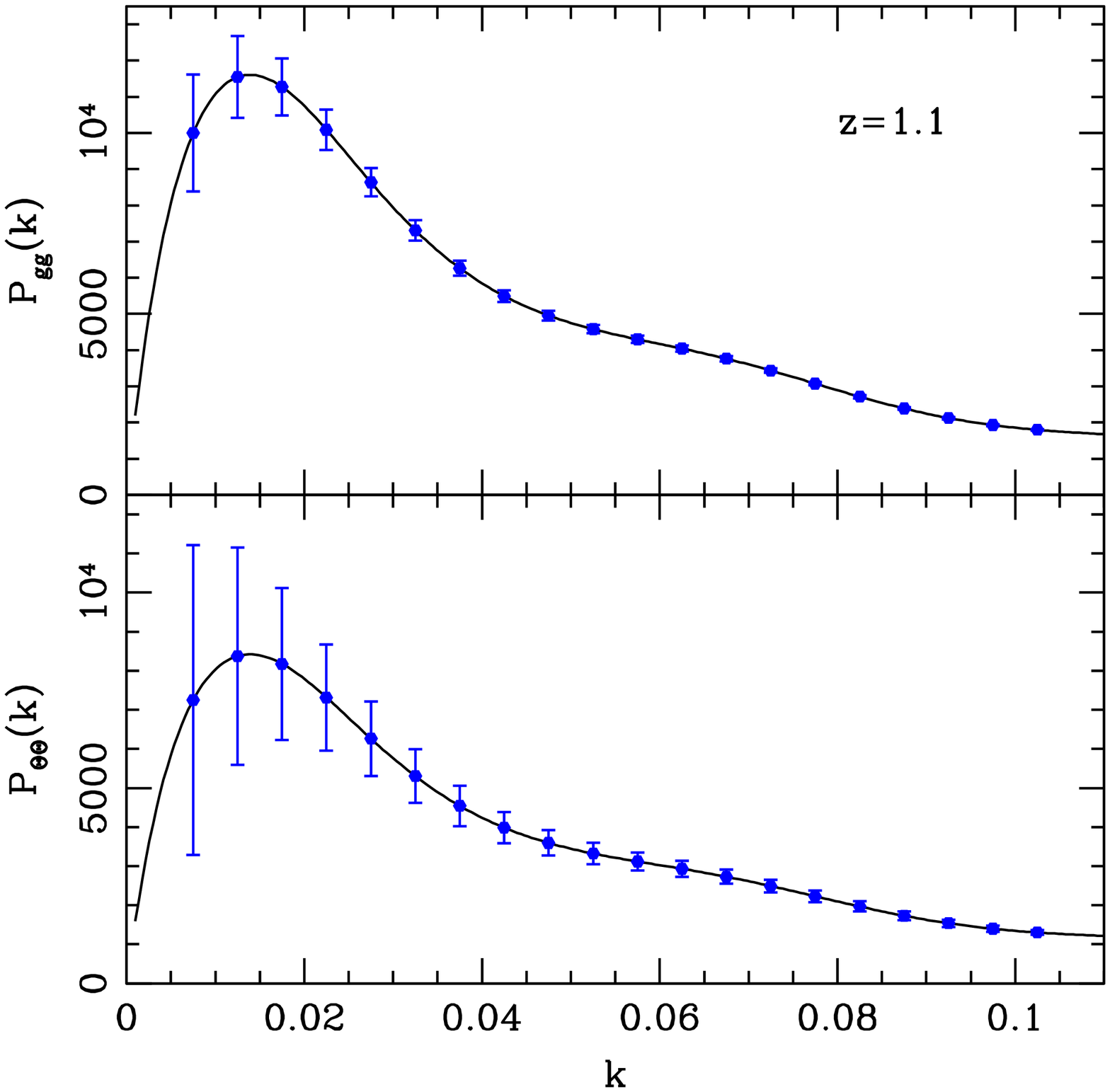}\hfill
\includegraphics[width=0.47\textwidth]{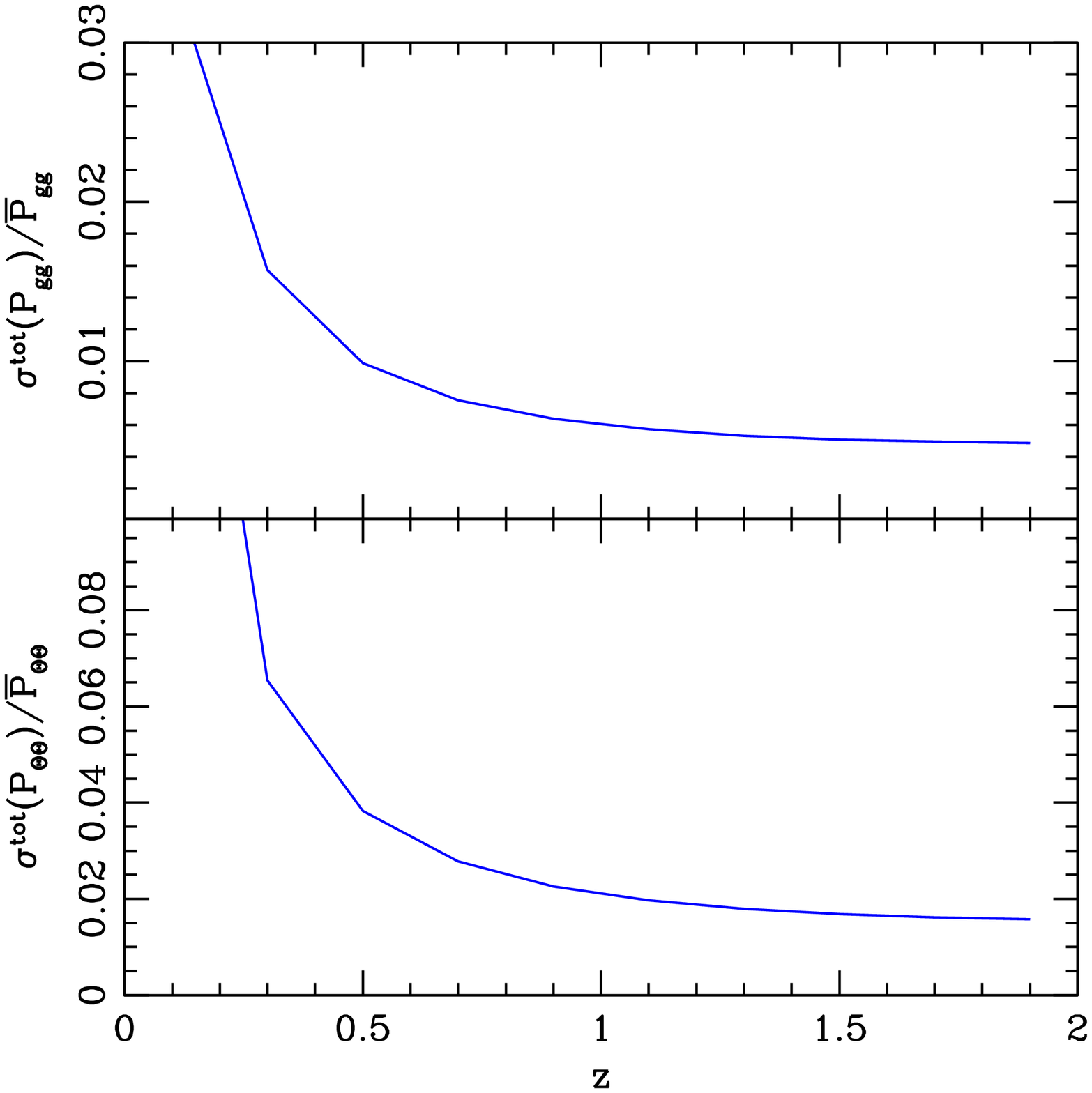}
\caption{\footnotesize {\it (Left panel)} Spectra at $z=1.1$ are presented. Solid curve in the top panel represents fiducial $P_{gg}(k)$, and solid curve in the bottom panel represents $P_{\Theta\Theta}(k)$. Error bars on the curves represent theoretical estimation using FLRW prior. {\it (Right panel)} Summed fractional errors of spectra at each given redshift are presented with applying FLRW prior.}
\label{fig:sigma_Friedman}
\end{figure*}

We split Fourier space into 20 bins from 0.005 $\mpcoh$ to 0.105 $\mpcoh$. The full set of parameter space of $p_{\alpha}$ becomes $(P_{gg}(k_i,z_j)$, $P_{\Theta\Theta}(k_i,z_j)$, $D_A(z_j)$, $H^{-1}(z_j))$. The  subscript ``$i$'' denotes 20 $k$ bins, and the subscript ``$j$'' denotes 10 $z$ bins from 0 to 2. The derivatives in terms of ($P_{gg}(k_i,z_j)$, $P_{\Theta\Theta}(k_i,z_j)$) in the Fisher matrix are given by,
\begin{eqnarray}\label{eq:dpdpk}
  \frac{\partial \ln \tilde{P}(k,\mu,z_j)}{\partial P_{gg}^{\dec}(k_i,z_j)}
  &=& \frac{\Theta_i(k)}{\tilde{P}(k,\mu,z_j)}
  \left[1 + \mu^2
  \sqrt{\frac{P_{\Theta\Theta}(k_i,z_j)}{P_{gg}(k_i,z_j)}} 
  \right] \nonumber \\
  \frac{\partial\ln \tilde{P}(k_i,\mu,z_j)}{\partial P_{\Theta\Theta}^{\dec}(k_i,z_j)}
  &=&\frac{\mu^2\Theta_i(k)}{\tilde{P}(k_i,\mu,z_j)}
  \left[\sqrt{\frac{P_{gg}(k_i,z_j)}{P_{\Theta\Theta}(k_i,z_j)}}+\mu^2\right]\nn\,,\\
\end{eqnarray}
where $\Theta_i(k)$ denotes step function in which $\Theta_i(k)=1$ at $k$ belong to $k_i$ bin, otherwise $\Theta_i(k)=0$. The derivative terms of distance measures are the same as in Eq.~\ref{eq:dpdD}. We find that none of quantities are observable in this full variation.

We show that the spectra of $P_{gg}(k_i,z_j)$ and $P_{\Theta\Theta}(k_i,z_j)$ can be observable using Approach II. With the FLRW prior, the parameter set of $p_{\alpha}$ becomes $(P_{gg}(k_i,z_j)$, $P_{\Theta\Theta}(k_i,z_j)$, $H^{-1}(z_j))$. In the left panel of Fig.~\ref{fig:sigma_Friedman}, the detectability of $P_{gg}(k_i,z_j)$ and $P_{\Theta\Theta}(k_i,z_j)$ at $z_j=1.1$ is presented in the top and the bottom panels respectively. There is no much difference from the case in which the distance measures are known~\cite{White:2008jy}.

We continue our test at the other redshift bins. Here we present the results using cumulated error of $\sigma^{\rm tot}(P)$ which is given by,
\ba
\left[\sigma^{\rm tot}(P)\right]^{-2}=\sum_{i,j=i_{k_{\rm min}}}^{i_{k_{\rm max}}}  P(k_i)C^{-1}[P(k_i),P(k_j)]P(k_j)\,,
\ea
where $C^{-1}$ is the inverse of the covariance matrix calculated using the Fisher matrix. In the top and the bottom panels of~Fig.~\ref{fig:sigma_Friedman}, we present the cumulated dispersion of $\sigma^{\rm tot}(P_{gg})$ and $\sigma^{\rm tot}(P_{\Theta\Theta})$ from $z=0.1$ to $z=1.9$. Both spectra are observable in precision using the FLRW prior.

\section{Discussion}

The spectra of density and velocity perturbations and the distance measures are not simultaneously measured due to high degeneracy among those observables. In this manuscript, two conditions are presented that this degeneracy is broken.

First, it can be resolved by a prior information of the shape of spectra. The spectra of density and velocity perturbations are composed of the shape and the coherent growth factors. The broadband shape can be parameterized by a few shape cosmological parameters of ($\omega_m$, $\omega_b$, $n_S$). The primordial shape during inflation is determined by $n_S$, and the transferred shape before the matter--radiation equality epoch is described by $\omega_m$ and $\omega_b$. The coherent growth factors of spectra are parameterized by $G_{g}$ and $G_{\Theta}$. We show that the growth functions of $G_{g}$ and $G_{\Theta}$ and the distance measures of $D_A$ and $H^{-1}$ are simultaneously observable in precision, when the shape parameters are marginalized with CMB experiments. 

Second, it can be resolved by assuming a minimal theoretical ansatz of the FLRW universe prior in which two distinct components of distance measures are unified into a single degree of freedom. The angular diameter distance of $D_A$ is expressed as an integrated sum of Hubble flow of $H^{-1}$. In practice, this reduction is possible by demanding that an experimental design of wide--deep field redshift survey be fully tomographic through all redshifts. Then the spectra and the distance measures are parameterized by ($P_{gg}$, $P_{\Theta\Theta}$, $H^{-1}$) without the angular diameter distance of $D_A$. We find that the spectra of $P_{gg}$ and $P_{\Theta\Theta}$ are measured in precision about a sub--percentage level. Additionally, it is remarkable that the Hubble flow of $H^{-1}$ is measured in high precision at all redshifts. The evolution of Hubble flow is a key observable to reconstruct the history of cosmic expansion. We are able to probe it in cosmological model independent way using this FLRW prior approach. The detectability of ($P_{gg}$, $P_{\Theta\Theta}$, $H^{-1}$) is minimally influenced by relaxing flat prior with assuming small curvature approximation.

However, we have to mention caveats. As it is well known, a redshift distortion map is severely contaminated by the non--linear smearing effect. It is indeed troublesome to decompose linear information due to this contamination. But many reports have been made of the possible precise decomposition of the linear spectra of density and velocity fluctuations using the `beyond Kaiser model'~\cite{Scoccimarro:2004tg,Matsubara:2007wj,Crocce:2007dt,Taruya:2010mx}. In near future, we will be back to the work to decompose the spectra and distance measures using simulated maps.

\acknowledgements{We thank Hee-Jong Seo for assistance to initiate this work and for substantial inputs in this manuscript. We thank Eric Linder for helpful discussions during a KITPC workshop. Numerical calculations were performed by using a high performance computing cluster in the Korea Astronomy and Space Science Institute.}

\end{document}